\documentclass[preprint,nofootinbib,tightenlines,amsmath,amssymb,showpacs]{revtex4}
\usepackage{graphicx}
\usepackage{dcolumn}
\usepackage{bm}
\usepackage{color}
\usepackage{amsfonts}
\begin{document}
\baselineskip=15pt \parskip=3pt

\vspace*{3em}

\title{Gauge theory of quantum gravity}

\author{Cao H. Nam}
\email{chnam@iop.vast.ac.vn} \affiliation{Institute of Physics,
Vietnam Academy of Science and Technology,\\  10 Dao Tan, Ba Dinh,
Hanoi, Vietnam}
\date{\today}

\begin{abstract}
The gravity is classically formulated as the geometric curvature
of the space-time in general relativity which is completely
different from the other well-known physical forces. Since seeking
a quantum framework for the gravity is a great challenge in
physics. Here we present an alternative construction of quantum
gravity in which the quantum gravitational degrees of freedom are
described by the non-Abelian gauge fields characterizing
topological non-triviality of the space-time. The quantum dynamics
of the space-time thus corresponds to the superposition of the
distinct topological states. Its unitary time evolution is
described by the path integral approach. This result will also be
suggested to solve some major problems in physics of the black
holes.
\end{abstract}


\maketitle

Developing of a consistent quantum theory of the gravity at very
small distances or high energy scales, near the Planck scale
$l_\textrm{P}\approx 10^{-35}$ m, has been interested by many
physicists in which the gravity is strongly played by the rules of
the quantum mechanics. In fact, a theory of the gravity is to
study the space-time. Since well understanding of the quantum
gravity allows us to deal with the behavior of the space-time at
very small distances. The main motivations for seeking this are
just purely theoretical because of the fact that there have not
had immediately experimental observations for the quantum
gravitational effects so far. However, some indirect experimental
constraints for Planck-length quantum properties of the space-time
have been exploited with a growing number of studies
\cite{Camelia1998,Camelia1999,Alfaro2000,Jacobson2003,Myers2003,Camelia2004,Jacob2007,Contaldi2010,Botermann2014}
(see also Ref. \cite{Hossenfelder2011} and references therein).
The classical gravity has failed in explaining some the important
questions in the black holes and the Big Bang. The present of
singularities, i.e. breakdowns of the predictability of the
physical equations, in which the curvature of the space-time and
energy densities become infinite, is unavoidable under general
conditions \cite{Hawking1973}. The ultraviolet divergences arise
as attempting treatment of general relativity as the quantum field
theory leading to the perturbative non-renormalization. Some other
important open issues can also be found in \cite{Kiefer2005}.
These have suggested that the gravity description of general
relativity is incomplete since a more comprehensive theory is
needed. Constructing such a theory definitely needs to a complete
picture of the quantum cosmology which allows us to know that what
happens in the beginning moments of the universe.

There have been many different approaches to quantize the
gravitation interaction. The well-known main ones consist of
quantum general relativity which is the application of the
quantization rules to general relativity \cite{Rovelli2006}, or
string theory. However, a full quantum theory of gravity seems
quite clear to have not yet been available. Thus the problem of
quantum gravity remains one of the most important tasks of high
energy physics which would be expected to come the next a few
decades.

At present, the immediate search for the quantum gravity effects
are quite distant to hope due to the fact that the Planck scale is
about $10^{14}$ times larger than working energy of near future
accelerators. It thus makes the major obstacle in constructing
quantum gravity. However, it is interesting in the
higher-dimensional universe that the quantum gravitational effects
become very relevant at a energy scale which could be much lower
than the traditional value $l^{-1}_\textrm{P}$. It is thus
potentially within reach of the future accelerators. The
possibility of lowering the fundamental Planck scale of the order
of the electroweak scale was first proposed by Arkani-Hamed,
Dimopoulos and Dvali \cite{Hamed1998,Antoniadis1998,Type1998}. For
earlier ideas the readers can see Ref. \cite{Antoniadis1990}.

As well known, most of the fundamental interactions in Nature
based on the gauge theory have been constructed to be physically
sensible due to that they lead to the consistent quantum
description. The gauge theories are in principle applicable up to
arbitrarily high energy scales. Thus, it would be natural to seek
a gauge theory structure for the gravitational interaction in
which the general relativity is derived as the low energy limit.
In Ref. \cite{MacDowell1977}, the gravity can be interpreted as a
gauge theory by gauging the Poincar\'{e} group of the tangent
space of the space-time. It is important to see that this
formulation is precisely dependent of the background metric
structure of the space-time where the Riemann curvature tensor can
be understood as the field strength of the corresponding gauge
fields. This tells the equivalence between the Poincar\'{e} gauge
theory of the gravity and Einstein gravity plus the non-vanishing
of torsion. The gauge theory formulation of the gravity which will
be given in later is not followed in the gauge theory of
(pseudo-)Riemannian manifolds.

In this paper, we provide a quantum description of the gravity
based on the braneless higher-dimensional space-time framework in
the natural units $\hbar=c=1$. The path integral techniques of the
quantum gauge theories will be used to quantize the gravity. In
the following, we would like to briefly remind the space-time of
interestingly geometric structure which was suggested in our
recent work \cite{Nam2014} where the analysis is provided in more
much detail.

It was considered as a $D$-dimensional manifold $B^D$ always
foliated by disjoint $(D - 4)$- dimensional submanifolds of which
are smoothly equivalent to a Lie group $G$, either
$\mathrm{SU}(N)$ or $\mathrm{SO}(N)$ group, but are not in general
$G$ themselves, called the internal spaces \cite {otheLiegr}. On
the other hand, there is an attached smooth copy of $G$ on each
point of an another $4$-dimensional pseudo-Riemannian manifold
with the Lorentz metric adopted the metric signature $(+,-,-,-)$.
Thus, the space-time should be naturally described in term of the
corresponding principal bundle. In order to develop physically
interesting results, we need to define the coordinates for each
point on $B^D$ which are in general given by
\begin{eqnarray}
X^M&=&x^\mu, \hspace*{0.5cm} M=0,1,2,3, \nonumber\\
X^M&=&\frac{\theta^a}{\Lambda}, \hspace*{0.5cm} M=4,...,D-1,
\end{eqnarray}
only on any local neighbourhood of $B^D$. The $4$-dimensional
external indices are labeled by $\mu,\nu,...=0,1,2,3$, and the
$(D-4)$-dimensional internal indices are associated with $a,b,...$
running from $1$ to $D-4$. A dimensionless set $\{\theta^a\}$
parameterizes every element of the Lie group $G$ through the
exponential map as, $g=\textrm{exp}\{i\theta^aT_a\}$, where $T_a$,
with $a=1,...,D-4$, are the generators of the Lie algebra of the
Lie group $G$. It is very important that each internal space is
obviously of the dimensionless manifold since an energy scale
$\Lambda$ occurs to characterize physically on it. One should not
confuse this scale with the inverse radii of the compact internal
spaces which is in fact determined only if their metric is
endowed. The local coordinate transformation is as a result of the
non-empty overlap of any two local neighbourhoods leading to the
new symmetry principles for the space-time read
\begin{equation}
x'^\mu=\Lambda(x){^\mu}_\nu x^\nu,\
e^{i\theta'^aT_a}=h(x)e^{i\theta^aT_a},\label{gct}
\end{equation}
where $\Lambda(x)$ and $h(x)=\exp\{i\alpha^a(x)T_a\}$ are elements
of the linear transformation group $GL(4,\mathbb{R})$ and the Lie
group $G$, respectively.

An interestingly main result of our proposed space-time is that
the internal spaces are split up in the independent way to the
$4$-dimensional external submanifolds of $B^D$ which are
transversal to the internal spaces and identified as the usual
observed $4$-dimensional world in topologically non-trivial
structure of the space-time. The directions along the internal
spaces are fully determined by the well-defined local frames
$\{\Lambda\partial_a\equiv\hat{\partial}_a\}$ that are for the
whole tangent space of the internal spaces. The directions along
the $4$-dimensional external spaces are impossible to define
because of depending on the non-triviality of the principal bundle
$B^D$. However, it was argued that there exist the non-Abelian
gauge fields $A_\mu(x)=A^a_\mu(x)T_a$ living on $B^D$ to give a
principled way of how to move from an internal space to another.
According to these fields, the local frames
$\{\hat{\partial}_\mu\equiv\partial_\mu-g_{\textrm{i}}A^a_\mu\partial_a\}$,
with $g_{\textrm{i}}$ being the dimensionless gauge coupling
constant, well span the whole tangent space of the $4$-dimensional
external spaces. Since any particle propagating along the
$4$-dimensional external directions will be coupled to $A^a_\mu$.
It is interesting that the corresponding gauge charges are
generated dynamically through its dynamics in the internal spaces.
This shows universal nature of gravity which interacts with all
forms of energy presenting in the space-time as well as gives a
new insight into physics associated with extra dimensions. Under
the transformation (\ref{gct}), the gauge fields transform as
\begin{equation}
A'_\mu(x')=\frac{\partial x^\nu}{\partial
x'^\mu}\left[h(x)A_\nu(x)h(x)^{-1}
+\frac{1}{ig_{\textrm{i}}}h(x)\partial_\nu
h(x)^{-1}\right],\label{gautrnI}
\end{equation}
which is to combine of both the $4$-dimensional external
coordinate and the usual gauge transformations.

The fundamental variables allowing us to study the dynamics of the
gravity are given in terms of the gauge fields $A^a_\mu$ and the
background fields, $g_{\mu\nu}$ and $\gamma_{ij}$
($i,j=1,...,D-4$), which define the infinitesimal line elements on
the $4$-dimensional external and internal spaces, respectively
\cite{gloinfra}. The present of these metric fields points that
the principal bundle $B^D$ come with the additional
pseudo-Riemannian structure. All them are unified in the
non-trivial geometrical framework of the space-time. Note that
$\gamma_{ij}$ is positive definition and completely independent of
$g_{\mu\nu}$. As shown in \cite{Nam2014}, the internal metric
$\gamma_{ij}$ consists of the globally dynamic degrees of freedom.
This led to considerable simplification as
\begin{equation}
\gamma_{ij}=e^{\phi_i}\delta_{ij},\label{ExGMI}
\end{equation}
where $\phi_{D-4}=0$. In this way, $\phi_i$ ($i=1,...,D-5$) play
the role of the modulus fields. Furthermore, it was argued that
the modulus fields are all independent of the internal coordinates
to guarantee the metric on each internal space to be invariant
under the action of $G$. The classically pure gravity action is
taken the form
\begin{equation}
S_{\textrm{gra}}=\int_{B^D}
d^4xd^{D-4}\theta\sqrt{|G|}\left[-\left(\frac{M_*}{\Lambda}\right)^{D-4}\frac{F^a_{\mu\nu}F^{\mu\nu
a}}{4}+\overline{M}^{2}\left(
R-V(\phi_i)\right)\right].\label{bgac}
\end{equation}
The fundamental Planck scale $M_{*}$ determines the working energy
of the quantum gravity, and
$\overline{M}^{2}=M^{D-2}_{*}/\Lambda^{D-4}$. $G$ is referred to
the determinant of the higher-dimensional metric. $F^a_{\mu\nu}$
is the usual Yang-Mills field strength tensor of $A^a_\mu$ which
in this structure measure the non-triviality of the principal
bundle space-time $B^D$, or its topological characters. $R$ is the
scalar curvature of the space-time defined to correspond with of
the Levi-Civita connection. It defines the geometric curvature of
the space-time where its topology has been fixed by the Yang-Mills
curvature $F^a_{\mu\nu}$. This term includes the conventional
$4$-dimensional Einstein-Hilbert term and the internal dynamics
for $g_{\mu\nu}$, the dynamical term of the modulus fields and the
coupling between them to the gauge fields $A^a_\mu$. $V(\phi_i)$
is an extension of $\phi_i$-dependence of bulk cosmological
constant which generate the potential of the modulus fields.
However, this potential would obtain other contributions coming
from the bulk determinant and the scalar curvature $R$ to provide
a full potential of the moduli stabilization. It is emerged as one
of the crucial natures of the space-time $B^D$. Thus, the physical
size of the internal spaces is completely fixed, and the
excitations of the modulus fields are thus massive. The readers
can see the explicit form of the last two terms in action above in
\cite{Nam2014}. The factor $(M_*/\Lambda)^{D-4}$ can be absorbed
by normalizing the gauge fields $A^a_\mu$ in which the new gauge
coupling constant,
$g_{\textrm{i}}\left(\Lambda/M_*\right)^{\frac{D-4}{2}}$, to
remain dimensionless. This shows that the coupling between
$A^a_\mu$ and other fields depends not only of the usual gauge
coupling constant $g_{\textrm{i}}$, but of the characteristic
energy scale $\Lambda$ of the internal spaces, the fundamental
Planck scale $M_{*}$ and the extra dimensions as well. It is
obviously as a result of the fact that these gauge fields connect
the external and internal spaces in the unified form of the
space-time.

Let us now treat of the quantum mechanics of the space-time by the
path integral quantization which is the manifestly covariant
formulation. This is expressed by vacuum-to-vacuum amplitude of
transition from an initial ($D-1$)-space state $t_i\rightarrow
-\infty$ to a final ($D-1$)-space state $t_f\rightarrow +\infty$
fixed all that will be evaluated in terms of sums over all
possible gravitational field configurations connecting these two
states weighted by an appropriative action. However, in order to
well obtain a path integral description of the gravity, we should
first know what is the precise meaning of this transition
amplitude. This is only achieved if all possible field
configurations for the quantum gravity are defined. In our present
work, we assume that these are given to correspond with the
distinct topological configurations of the space-time. This means
that the sum over the paths is taken on the configuration space of
all space-times without distinguishing smooth structure, or
background metric, equipped on it. In this way, the configurations
are classifiable with respect to the proposed space-time through
all possible configurations of the gauge fields $A^a_\mu$
unrelated by the gauge transformations (\ref{gautrnI}). This is
dealt with by the choice of an appropriative gauge fixing term
added to the original gravitational Lagrangian to pick out an
unique element from each gauge equivalence class. The structure of
the space-time at small distances thus is a quantum-mechanical
superposition of the distinct topological states dynamically
generated from the fundamental degrees of freedom $A^a_\mu$
playing the role of quanta with respect to the space-time. It is
also important that the macroscopic background structure of the
space-time is fixed by the classically geometric dynamical
variables of the external metric $g_{\mu\nu}$ and the modulus
fields $\phi_i$. Note that, these classical fields given in the
present framework are to correspond with the fundamental degrees
of freedom rather than the effective ones as occurring in the
gauge theory constructions for the gravitational interaction
\cite{Lasenby1998}. Therefore, the quantum fields of the gravity
are given by the non-Abelian gauge fields $A^a_\mu$ leading to the
path integral variables whereas both $g_{\mu\nu}$ and $\phi_i$ are
realized as the $c$-number quantities or the classical fields
under consideration.

The true dynamical information of the quantum gravity is contained
in the vacuum-to-vacuum transition amplitude that is taken the
form in the presence of sources as
\begin{equation}
\langle 0|0\rangle_{J,\bar{\eta},\eta}= Z[J,\bar{\eta},\eta]=N\int
\mathcal{D}A_\mu\mathcal{D}\bar{c}\mathcal{D}ce^{iS^{(J,\bar{\eta},\eta)}_{\textrm{tot}}}\label{Gpathint}.
\end{equation}
The factor $N$ is a normalization constant so that $\langle
0|0\rangle=1$ in which all sources are absent. The Grassmann
fields $c$ and $\bar{c}$ are known as the ghost and anti-ghost
fields, respectively, which are both to depend of the only
$4$-dimensional external coordinates. The total gravity action
$S^{(J,\bar{\eta},\eta)}_{\textrm{tot}}$ is obtained by adding to
the original gravitational action a gauge fixing action and a
Faddeev-Popov ghost action as well as sources read
\begin{equation}
S^{(J,\bar{\eta},\eta)}_{\textrm{tot}}=S_{\textrm{gra}}+\left(\frac{M_*}{\Lambda}\right)^{D-4}\int_{B^D}
d^4xd^{D-4}\theta\sqrt{|G|}\left[-\frac{1}{2\xi}(\nabla^\mu
A^a_\mu)^2-\bar{c}^a\nabla^\mu(D_\mu c)^a+J^\mu
A_\mu+\bar{\eta}c+\bar{c}\eta\right],
\end{equation}
where $\xi$ represents an arbitrary gauge fixing parameter, and
\begin{eqnarray}
\nabla^\mu X^a_\mu&=&g^{\mu\nu}(\partial_\mu
X^a_\nu-\Gamma^\rho_{\mu\nu}X^a_\rho), \\
(D_\mu c)^a&=&\partial_\mu c^a-g_{\textrm{i}}f^a_{bc}A^b_\mu c^c,
\end{eqnarray}
here $X^a_\mu$ is referred to $A^a_\mu$ and $(D_\mu c)^a$, and
$\Gamma^\rho_{\mu\nu}$ are the $4$-dimensional external
coefficients of the Levi-Civita connection. The gauge fixing
breaks the usual gauge invariance but remain the $4$-dimensional
external covariance. The Faddeev-Popov ghost appear to subtract
out the unphysical degrees of freedom coming from the gauge fields
$A^a_\mu$. $J^\mu$ is interpreted as a source coupled to the gauge
fields $A^a_\mu$ while $\eta$ and $\bar{\eta}$ are the
Grassmannian sources for $c$ and $\bar{c}$, respectively. As
mentioned above, the background metric described by the classical
fields since the quantum-mechanical transition amplitude above has
not been integrated over all geometries of the space-time.

The generating functional $Z[J,\bar{\eta},\eta]$ encodes
everything about the quantum dynamics since evaluating it will
allow to derive results of the quantum gravity phenomenology
without the conceptual difficulties. In particular, it is
straightforward to include non-gravitational quantum fields. On
the other hand, the quantization of the gravity by
(\ref{Gpathint}) provides a better quantitative understanding of
quantum gravity to elucidate the topological dynamic nature of the
quantum space-time. In the weak coupling low energy region,
$Z[J,\bar{\eta},\eta]$ can be computed by using the perturbation
expansion in classical $4$D-Minkowski background which leads to
the Feynman diagrams of scattering amplitude containing the
gravitational quantum fields $A^a_\mu$. Within this case, the
ultraviolet behavior of the quantum gravity arising from the high
momenta contributions in the loop corrections will be well defined
by the renormalization of gauge theories. On the other hand, this
quantization will be devoid of the ultraviolet infinity problem of
quantum gravity making the physically relevant predictions at low
energies. More importantly, the non-perturbative effects of the
quantum gravity expected by reasons play significantly important
roles at the extremely small distances or the extremely high
energy scales. Since investigations of the strongly coupled regime
of quantum gravity requires the use of non-perturbative methods to
define a regularized form of the above generating functional.
Therefore, within the present framework the quantum gravity is in
principle well behaved at any energy scale.

It is important to note here that the proper description of the
space-time is dealt with the consistent hybrid dynamics: the
quantum one is for describing the very small-scale structure of
the space-time and the classical one is for doing the large-scale
behaviour of the space-time. It is interesting in obtaining these
that are performed via the fundamentally dynamical variables. We
have already seen that the quantum properties of the space-time is
full given by the quantum generating functional above. The
classical properties of the space-time can completely defined by
the equations of motion for the fields corresponding with the
integral in (\ref{Gpathint}) to be invariant under the slight
translation shift of the fields
\begin{equation}
\delta_\phi\langle 0|0\rangle=iN\int
\mathcal{D}A_\mu\mathcal{D}\bar{c}\mathcal{D}c\delta_\phi
Se^{iS}=i\langle 0|\delta_\phi S|0\rangle Z\equiv i\langle
\delta_\phi S\rangle Z,
\end{equation}
where $\phi$ stands for the above given gravitational fields. This
is well known as the Schwinger's action principle. The
corresponding equations of motion for the external metric field
$g_{\mu\nu}$ and the mass-dimensional normalized modulus fields
$\widetilde{\phi}_i=\overline{M}\phi_i$ can be written in the form
\begin{eqnarray}
G_{\mu\nu}+...&=&\overline{M}^{-2}\left[T_{\mu\nu}(\widetilde{\phi}_i)+\langle
T_{\mu\nu}(A^a_\mu,\bar{c}^a,c^a)\rangle+\langle T_{\mu\nu}(J^\mu,\bar{\eta},\eta)\rangle\right],\label{Einseqs}\\
\sqrt{|G|}^{-1}\hat{\partial}_\mu(\sqrt{|G|}\hat{\partial}^\mu\widetilde{\phi}_{i})&=&J_i(g_{\mu\nu},\widetilde{\phi}_{i})+\langle
J_i(A^a_\mu,\bar{c}^a,c^a)\rangle+\langle
J_i(J^\mu,\bar{\eta},\eta)\rangle,\label{modfieldeqs}
\end{eqnarray}
where $G_{\mu\nu}$ is Einstein tensor expressed in term of the
metric $g_{\mu\nu}$. The terms of the right-hand side of Eq.
(\ref{Einseqs}) and (\ref{modfieldeqs}) are realized as the
sources where the metric $g_{\mu\nu}$ and the modulus fields
$\widetilde{\phi}_{i}$ are both sourced by the quantum nature
fields through their vacuum expectation value (VEV). It is well to
understand that these can be expressed via basic macroscopic
parameters, such as total mass-energy, electric charge or angular
momentum, of source. The ellipsis contains terms determining the
dynamics of $g_{\mu\nu}$ in the internal spaces which would be not
interesting to explicitly write here. These field equations whose
two hand sides are equated by $c$-number quantities determines the
macroscopic geometry of the space-time. In other words, the
curvature and geometry of the space-time is defined in which its
classically induced topology is replicated via the quantum
averaging on all quantum topological excitations as
\begin{equation}
\langle D_\mu F^{\mu\nu a}\rangle=-\langle J^{\nu
a}\rangle+(\emph{other sources}).\label{YMeq}
\end{equation}

As we have already seen that the quantum and classical
gravitational effects are independent but influenced together. The
microscopic properties of the space-time act on its large-scale
structure via the reaction terms appearing in the above classical
field equations in which the induced classical field description
is produced by the collective retroaction of a very large number
of quantum fields excited. The macroscopic properties acting on
its small-scale structure are expressed via their presence in the
vacuum-to-vacuum amplitude. However, we would emphasize that the
Eq. (\ref{Einseqs}) and (\ref{modfieldeqs}) are not thought as
those in the semiclassical approximation which was originally
suggested by M{\o}ller \cite{Moller1962} and Rosenfeld
\cite{Rosenfeld1963} in which they describe the coupling between
the classical gravitational system with a quantum system. It
argued that this coupling type leads to inconsistencies in
particular in the form the quantum decoherence with an
irreversible environment. The present construction is to be
distinguished from such type because of that the relation between
the quantum and classical effects is established without creating
contradictions in the unified manner of the same object of the
space-time. On the other hand, the classical gravitational fields
can not in principle be used to make a measurement on state of the
quantum fields leading to the quantum wave function collapse.
Since the present theoretical framework is able to provide a
sensible way to unify quantum mechanics and general relativity,
which have played the major roles in their own arenas, that both
are of the nature of the space-time.

We would like to emphasize that the quantization of the gravity
proposed in the present framework can overcome some difficulties
appearing in attempt to quantize the background metric field. The
background metric is described by the classical physical fields
meaning that it has a defined value since the causal structure of
the space-time is completely well-defined. This becomes
ill-defined when the metric is treated as the quantum fields
leading to an undefined value. An interestingly particular aspect
of the dynamics of the quantum matter on the quantum space-time is
that it propagates in a superposition of the distinct topological
quantum states expressed in coupling to $A^a_\mu$ whereas that on
the classical background of the space-time must follow a
superposition of either timelike or null world-lines. These result
in a interestingly physical consequence that time evolution with
an initial and a final states fixed is forbidden by the classical
causal structure, but is still possible to be allowed in the
quantum sense. For example, such evolution occurs in the black
hole as we will discuss in later. The study of background metric
quantization in a full manner is just very difficult due to the
the kinetic term of the metric field has no quadric form. Thus,
the metric would be usually expanded around a fixed background,
such as the Minkowski-flat metric, in which the small oscillation
about this is realized as the quantum field, known as the
covariant perturbation method. However, the main difficulty
arising in this framework is that the quantum gravity is not
perturbatively renormalizable. With the macroscopical metric, the
present work can allow to understand why we observe a classical
universe as a whole from a quantum universe at the very early
moments.

It can be checked that the total gravitational Lagrangian density
is invariant under global fermionic transformations known as the
BRST ones (the above gauge fixing term should be replaced by
another one including auxiliary fields which differs by a total
divergence) in which $\delta g_{\mu\nu}=0$ and
$\delta\widetilde{\phi}_i=0$ under these transformations. In
addition to these, the total gravitational Lagrangian density is
also invariant under global bosonic transformations known as the
ghost scaling symmetry given by
\begin{equation}
\delta c^a=\epsilon c^a,\  \ \delta\bar{c}^a=-\epsilon\bar{c}^a,
\end{equation}
where $\epsilon$ is a commuting infinitesimal constant while the
others are both inert under these. The N\"{o}ether charges
corresponding to these two types of symmetry denoted by $Q_{BRST}$
and $Q_c$ are conserved. Hence they can be used to define the
Hilbert space for the physical states of the quantum space-time
annihilated by them in the canonical operator quantization as
\begin{equation}
Q_{BRST}|\Psi_{\textrm{phys}}\rangle=0,\ \
Q_c|\Psi_{\textrm{phys}}\rangle=0.
\end{equation}
These conditions imply that the physical states should have the
zero ghost number which is sum of the number of the ghost and
anti-ghost particles. It is precisely due to that the ghost and
anti-ghost fields are unphysical fields. In this case, under the
unitary time evolution these states obey the time-dependent
Schr\"{o}dinger equation rather than Wheeler-DeWitt equation in
which the time parameter is related to a foliation of the
space-time by the ADM approach \cite{ADM}.

We extract now a special remark about the path integral
quantization given in above in which the quantum gravity theory
has an underlying topological symmetry. We have seen that each
path in the generating functional of the gravity above obviously
consists of much topologically equivalent space-times but may be
different smoothly. Since there exist the underlying topological
transformations $\{\delta_{ts}\}$ acting the quantum fields of
gravity which preserve the field configurations of the quantum
gravity leading to remain the generating functional
(\ref{Gpathint}). These symmetry ones are the symmetry of the
total action of gravity sastifying
$\delta_{ts}S^{(J,\bar{\eta},\eta)}_{\textrm{tot}}=0$. This
implies that the gravity quantization holds the existence of a set
of operators $\mathcal{O}_\alpha$ in the theory which are in
general arbitrary functions of the gauge fields $A^a_\mu$, the
ghost fields $c^a$ and the anti-ghost ones $\bar{c}^a$ such that
their correlation functions do not depend on the background
metric. Then the following relations are satisfied
\begin{equation}
\frac{\delta}{\delta
g_{\mu\nu}}\langle\mathcal{O}_1...\mathcal{O}_n\rangle=0,\  \
\frac{\delta}{\delta\phi_i}\langle\mathcal{O}_1...\mathcal{O}_n\rangle=0
\end{equation}
with $i=1,...,D-5$. These operators are thus called topological
observables. It is important to note that the symmetry
$\delta_{ts}$ is that of the quantum theory since it is not
anomalous. Therefore, one can say that the quantum field theory of
gravity may be topological, and thus predict the existence of the
new quantum topological observables.

In our discussion so far, we have studied the quantization of the
gravity in which the gauge fields $A^a_\mu$ are massless. This
means that the quantum effects should have been detected by
experiments which is clearly not consistent with the observed
fact. The gauge symmetry group $G$ has thus to be broken
spontaneously by non-zero VEV of a scalar field $\Phi$ that
rotates under the transformation (\ref{gct}) as
\begin{equation}
\Phi(x,\theta)\rightarrow\Phi'(x',\theta')=U\Phi(x,\theta),\label{PhiTr}
\end{equation}
where, $U=D[h(x)]$, is a matrix corresponding to a representation
of an element $h(x)\in G$ given in (\ref{gct}). A detail study to
such field is given in \cite{Nam2014}. Consequently, the gauge
fields $A^a_\mu$ get the masses from VEV of this scalar field
$\Phi$ to become massive. This leads to the quantum effects of the
space-time only play the important roles at the small distances or
the high energies. In the presence of spontaneous symmetry
breaking, the quantization of the gravity should be performed in
the Higgs phase in which the gauge fixing term must be changed as,
$(\nabla^\mu A^a_\mu-\xi M\Phi)^2/2\xi$, with $M$ being mass
matrix of $A^a_\mu$, which also induces modified ghost term. At
low energy region or large distance, the gauge fields and the
modulus fields are heavy which is why they have not been seen
experimentally while the external metric field are only remnant of
the full description reducing to general relativity. An effective
dynamics of the background metric fields below the
compactification scale in the weak field approximation is given in
Ref. \cite{Nam2014}.

We now have the space-time particularly respecting the principles
of quantum mechanics that can allow us for studying physics of the
black holes in a detailed quantitative picture \cite{quanBL} (as
well as in quantum cosmology). In extremely strong field and high
matter density near the black holes, quantum effects of the gauge
fields $A^a_\mu$ are strongly enough to break the classical smooth
structure of the space-time at small distances. This means that
the quantum effects of gravity will change the classical dynamics,
or more precisely the quantum matter dynamics on the classical
curved space-time, by the quantum dynamics in which the fields
propagate in the superposition of the quantum states of the
space-time. In this way, the quantum dynamics of the black holes
in the time corresponding to the formation and evaporation of
black holes is expressed by the generating functional
(\ref{Gpathint}) in which the initial and final states are assumed
to take the limit $t_f-t_i\rightarrow\infty$ while intermediate
states are shown by sources of the gravitational fields. Thus the
time evolution of this quantum mechanical system is a unitary
process. It will be convenient if we are interested in the
Euclidean approach performed by a Wick rotation leading to the
path integral representation of the partition function with
respect to an statistical ensemble. Since the statistical behavior
of the quantum black holes can be appropriately investigated via a
grand canonical ensemble at finite temperature
\cite{Gibbons-Hawking1976} in which a chemical potential term will
be included. An entropy of black hole corresponding to the
fundamental degrees of freedom underlying the quantum gravity
provides a microscopic explanation for the entropy of the black
hole, well-known as the Bekenstein-Hawking entropy due to emit
thermally the Hawking radiation. However, it should be noted that
this radiation arises as the result of the application of quantum
mechanics to electromagnetic fields on a classical curved
background space-time near a black hole where the quantum
properties of gravity are not investigated. Thus, there will have
extra energy fluxes beyond that predicted by Hawking'
semiclassical approximation calculation. On the other hand, black
hole entropy followed on this framework will be different to the
Bekenstein-Hawking entropy due to quantum corrections coming from
the space-time. It is of interesting to this framework that the
existence of the event horizon of the black hole arise from
thermodynamic averaging corresponding to the macroscopic geometry
of the space-time. This means that the quantum space-time do not
have the event horizons. In the quantum picture of the space-time
the fields must not follow the timelike and null paths in
classical viewpoint in which they would go to a future singularity
``\emph{point}'' \cite{singpoint} if they are inside the event
horizons. Since our work ensure that the information about the
quantum state of infalling matter is not destroyed by classical
singularity of the black holes. This means that the quantum
gravity effects prevent the information loss, breaking of quantum
correlation (or entanglement) as well as the energy density
blowing up to infinity at a place of the space-time to form the
classical curvature singularity.

We note here that the indirect manifestations of the space-time
quantization may be involved exchanges of virtual gauge fields
$A^a_\mu$ of high energies. For example, quantum gravitational
corrections ar possible to occur in the scattering process
$e^+e^-\rightarrow e^+e^-$, or contribute to muon $g-2$ induced
all by $A^a_\mu$. Besides, some constraints on quantum gravity
effects can be performed through the astrophysical bounds and
cosmology, such as from modified dispersion relations of
electromagnetic radiation coming from far-away objects by
microstructure of the space-time. Therefore, the quantum gravity
phenomenology of of this scenario at low energy is just very rich.

In summary, we have shown that the quantum nature of the
gravitational field or the space-time is well understood through
the quantum dynamics of the non-Abelian gauge fields by the path
integral method. Its classical nature is describe by the
background metric formulated in general relativity. The
quantization following this theoretical framework requires the
existence of the extra dimensions forming the internal spaces of
the space-time. These gauge fields provide a way to distinguish
the topological properties of the space-time through their
Yang-Mill curvature tensor. This allows us to make the
quantization of gravity in a quantum-mechanical superposition of
the distinct topological states.



\begin{thebibliography}{99}

\bibitem{Camelia1998} G. Amelino-Camelia, J. Ellis, N. E. Mavromatos, D. V.
Nanopoulos, and S. Sarkar, Nature {\bf 393}, 763 (1998).

\bibitem{Camelia1999} G. Amelino-Camelia, Nature {\bf 398}, 216 (1999).

\bibitem{Alfaro2000} J. Alfaro, H. A. Morales-Tecotl, and L. F. Urrutia,
Phys. Rev. Lett. {\bf 84}, 2318 (2000).

\bibitem{Jacobson2003} T. Jacobson, S. Liberati, and D. Mattingly, Nature {\bf
424}, 1019 (2003).

\bibitem{Myers2003} R. C. Myers and M. Pospelov, Phys. Rev. Lett. {\bf 90},
211601 (2003).

\bibitem{Camelia2004} G. Amelino-Camelia and C. L\"{a}mmerzahl, Class. Quantum Grav. {\bf 21}, 899
(2004).

\bibitem{Jacob2007} U. Jacob and T. Piran, Nature Physics {\bf 3}, 87
(2007).

\bibitem{Contaldi2010} C. R. Contaldi, F. Dowker, and L. Philpott, Class. Quant. Grav. {\bf 27}, 172001
(2010).

\bibitem{Botermann2014} B. Botermann \emph{et al}., Phys. Rev. Lett. {\bf 113}, 120405
(2014).

\bibitem{Hossenfelder2011} S. Hossenfelder, Classical and Quantum Gravity: Theory, Analysis and Applications, edited by V. R. Frignanni, Nova Publishers
(2011) arXiv:1010.3420 [gr-qc].

\bibitem{Hawking1973} S. W. Hawking and G. F. R. Ellis, \emph{The large scale structure of
space-time}, Cambridge University Press, Cambridge (1973).

\bibitem{Kiefer2005} C. Kiefer, Annalen Phys. {\bf 15}, 129
(2005).

\bibitem{Rovelli2006} C. Rovelli, \emph{Quantum Gravity}, Cambridge University
Press, Cambridge (2006).

\bibitem{Hamed1998} N. Arkani-Hamed, S. Dimopoulos, and G. Dvali, Phys.
Lett. B {\bf 429}, 263 (1998).

\bibitem{Antoniadis1998} I. Antoniadis, N. Arkani-Hamed, S. Dimopoulos, and G. Dvali,
Phys. Lett. B {\bf 436}, 257 (1998).

\bibitem{Type1998} G. Shiu and S.-H. Type, Phys. Rev. D {\bf 58}, 106007 (1998).

\bibitem{Antoniadis1990} I. Antoniadis, Phys. Lett. B {\bf 246}, 377
(1990).

\bibitem{MacDowell1977} S. W. MacDowell and F. Mansouri, Phys. Rev. Lett. {\bf 38}, 739 (1977) [Erratum {\bf 38}, 1376 (1977)].
The earliest work with the Lorentz group is, R. Utiyama, Phys.
Rev. {\bf 101}, 1597 (1956). Some recently interesting results
based on this object can be seen, for example, J. Lee, John J. Oh,
and H. S. Yang, JHEP {\bf 1112}, 025 (2011); H. S. Yang, PoS
CORFU2011 (2011) 063; J. J. Oh and H. S. Yang, Mod. Phys. Lett. A
{\bf 28}, 1350097 (2013).

\bibitem{Nam2014} C. H. Nam, arXiv:1407.8493 [hep-th].

\bibitem{otheLiegr} Other connected compact Lie groups are possible to be
generalized from Ref. \cite{Nam2014} without changing much.

\bibitem{gloinfra} In this case, the $(D-4)$-dimensional internal
indices are labelled to associate with $i,j,...$ to refer that the
internal metric is written in term of the global internal coframe
due to the group Lie $G$ acting smoothly and freely on $B^D$ which
is transitive on each internal space.

\bibitem{Lasenby1998} A. Lasenby, C. Doran, and S. Gull, Phil. Trans. Roy. Soc. Lond. A {\bf 356}, 487
(1998).

\bibitem{Moller1962} C. Moller, \emph{Les Theories Relativistes de la Gravitation}
Colloques Internationaux CNRX 91 ed A Lichnerowicz and M-A
Tonnelat (Paris: CNRS) (1962).

\bibitem{Rosenfeld1963} L. Rosenfeld, Nucl. Phys. {\bf 40}, 353 (1963).

\bibitem{ADM} R. Arnowitt, S. Deser, and C.W. Misner, ``\emph{The dynamics of general relativity}'',
in \emph{Gravitation: An introduction to current research}, edited
by L. Witten, Wiley (1962).

\bibitem{quanBL} C. H. Nam, in preparation.

\bibitem{Gibbons-Hawking1976} G. W. Gibbons and S. Hawking, Phys.
Rev. D {\bf 15}, 2752 (1976).

\bibitem{singpoint} Within the given structure of the space-time,
a classical singularity of black hole will occur to corresponding
with geometry of the $4$-dimensional external spaces. Thus this
singularity is precisely not a point but be an internal space.

\end{thebibliography}
\end{document}